\documentclass [12pt]{article}
\usepackage{graphicx,amssymb,amsfonts,latexsym,amsmath,amsthm,times}
\usepackage{epsfig}
\usepackage{fancyhdr}
\usepackage{color}
\usepackage[english]{babel}
\numberwithin{equation}{section}

\begin{document}

\footnotesize {\flushleft \mbox{\bf \textit{Math. Model. Nat.
Phenom.}}}
 \\
\mbox{\textit{{\bf Vol. XX, No. XX, 2012, pp. XX--XX}}}

\thispagestyle{plain}

\vspace*{2cm} \normalsize \centerline{\Large \bf Simple Model of Shape Evolution of Desiccated Colloidal Sessile Drop}

\vspace*{1cm}

\centerline{\bf Yu. Yu. Tarasevich\footnote{Corresponding
author. E-mail: tarasevich@aspu.ru}, I. V. Vodolazskaya,
 and O. P. Isakova}

\vspace*{0.5cm}

\centerline{Institute of Physics and Mathematics, Astrakhan State
University, 414056 Astrakhan, Russia}


\vspace*{1cm}

\noindent {\bf Abstract.}
We propose simple model of colloidal sessile drop desiccation. The model describes correctly both evolution of the phase boundary between sol and gel inside such a drop and the final shape of the dried film (deposit). The model is based on mass conservation and natural assumption that deposit (gel phase) prevents flows and evaporation.
\vspace*{0.5cm}




\vspace*{1cm}

\setcounter{equation}{0}
\section{Motivation}\label{sec;motivation}

During last decade, the drying of particle-laden sessile drops attracts much attention from the scientific community because of numerous applications. First of all one should refer to the important surface-patterning
technique known as evaporation-induced self-assembly (EISA) or evaporation driven self-assembly (EDSA), ink-jet printing, coating technologies, spotting technologies for bio-assays, \emph{etc.}

As a rule, the particle-laden sessile drops desiccate with pinned contact line. The outward flow occurs inside the drop and carries the particles to the drop edge.  The ring-like deposit remains on the substrate~\cite{Deegan2000,Deegan1997}.

There may be two very different situations when a particle-laden sessile drop desiccates.
First, the particles inside such a drop may interact with each other only pure mechanically (impacts). In this case, the deposit forms a porous medium. This medium prevents neither bulk flow inside it nor evaporation from its free surface.
Second, the particles may form chemical bonds. In the last case, a phase transition may occur (\emph{e.g.}, sol to gel). The new phase blocks any flows, the solvent evaporation from such a phase is negligible.
The proposed models mainly deal with the first situation~\cite{Attinger2009,Sefiane2009Langmuir,Deegan2000,Fischer2002,Kim2010,Kistovich2010,Ozawa2005,Parisse1996,Petsi2010,Popov2005,Widjaja2008,Witten2009,Zheng2009}.   Only several models treat the deposit as impenetrable for flows and inhibiting evaporation~\cite{Okuzono2009,Ozawa2005,Tarasevich2011CPS,Vodolazskaya2011MPLB}. Nevertheless, the simulation of desiccated colloidal drops with phase transition is extremely important for high-throughput drag screening~\cite{Takhistov2002}, biostabilization~\cite{Ragoonanan2008}, and medical tests~\cite{Rapis2002,Savina1999,Shabalin2001}. The models~\cite{Okuzono2009,Ozawa2005,Tarasevich2011CPS,Vodolazskaya2011MPLB} utilize the sets of rather complicated PDE's. We propose alternative very simple but efficient model.

\begin{figure}[htbp]
\centerline{\includegraphics[scale=0.5]{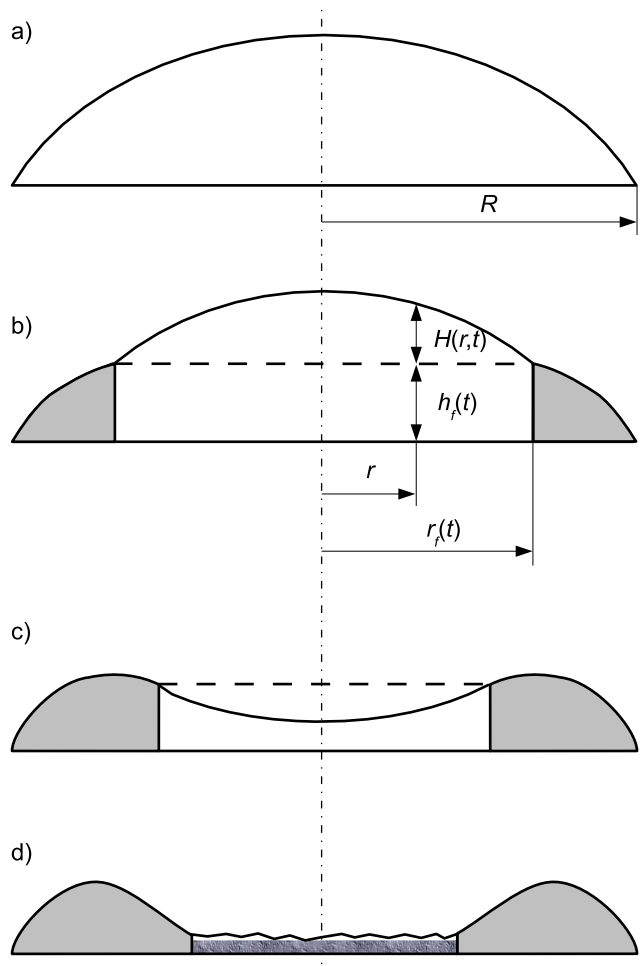}}
\caption{Drop evolution. Sol is shown in white. Gel is shown in gray.}
 \label{fig:dropevolution}
\end{figure}

There are three different modes of colloidal sessile drop desiccation~\cite{Pauchard1999}:
\begin{enumerate}
  \item $t_\text{G}\gg t_\text{D}$, where $t_\text{G}$ is the gelation time and $t_\text{D}$ is the desiccation time. The gelled phase occurs near the drop edge and moves inward while the central area of the drop remains liquid.
  \item $t_\text{G}\approx t_\text{D}$. The gelled skin covers the free drop surfaces. This thin shell cannot prevent evaporation of the solvent. The buckling instability occurs.
  \item $t_\text{G}\ll t_\text{D}$. The phase transition from sol to gel in the whole bulk of the drop is almost instantaneous. The gelled drop loses solvent via evaporation very slow.
\end{enumerate}
The mode of our interest is the first one. In this particular case, the desiccation process can be divided into several stages.
\begin{enumerate}
  \item Initial single-phase liquid stage (Figure \ref{fig:dropevolution}a). The whole drop is a sol. The outward flow carries suspended particles to the drop edge until volume fraction of the suspended particles, $\Phi$, reaches critical value, $\Phi_\text{g}$. Note that particle-enriched region is extremely narrow whereas the particle volume fraction in central area of the drop is almost constant along its radius~\cite{Tarasevich2010TP,Tarasevich2007JTP,Tarasevich2007epje}.
  \item Intermediate two-phase stage (Figure \ref{fig:dropevolution}b,c). Gelled ring near the drop edge is formed and growths to the drop center. The volume fraction of the colloidal particles is constant both inside the gelled foot, $\Phi_\text{g}$, and sol, $\Phi$, except rather narrow area near the phase  front~\cite{Tarasevich2011CPS,Vodolazskaya2011MPLB}.
  \item Final single-phase solid stage (Figure \ref{fig:dropevolution}d). The gelled deposit loses the rest of moisture and cracks. The real fluids of interests (\emph{e.g.}, biological fluids) may content both suspended particles and dissolved substances. In this case, the dendritic crystals may occur in the central area of a sample \cite{Annarelli2001,Takhistov2002}.
\end{enumerate}

In the present article, we focus our efforts on the second stage only.
The construction of this paper is as follows. In Section~\ref{sec:model}, we describe our model. In
Section~\ref{sec:results} the obtained results are discussed. Finally, we summarize our results and conclude this paper.

\vspace*{0.5cm}
\setcounter{equation}{0}
\section{Model description}\label{sec:model}
\subsection{Main assumptions}\label{sec:assumptions}

We consider a colloidal (nano- or microparticle-laden) sessile drop deposited on horizontal solid impermeable substrate. The drop is axisymmetric. In our consideration, we suppose
\begin{enumerate}
\item The liquid is incompressible.
\item The contact line is pinned during the desiccation.
\item Free surface of the liquid phase is a spherical cap. This assumption is valid for the sufficient small drops when Bond number is much smaller than unity and rather slow flows when capillary number is small.
\item Evaporation from the gelled phase is negligible and doesn't effect on hydrodynamics inside liquid phase. This assumption is supported by the experimental data~\cite{Okuzono2010JPSJ,Sobac2011PRE}.
\item Any hydrodynamical flows inside gelled phase are absent. This statement can be concluded from the simple observation that gelled ring cracks (see, \emph{e.g.}, \cite{Annarelli2001}, \emph{i.e.} there is not any flow to prevent moisture loss even by very slow evaporation.
\item The drop apex height as well the drop mass decrease almost linearly up to full gelation \cite{Annarelli2001}.
\item The phase boundary is vertical. Volume fraction of the colloidal particles is $\Phi(t)$ in the liquid (sol) phase and constant $\Phi_\text{g}$ in the solid (gel) phase. This assumption is supported by our simulations~\cite{Tarasevich2011CPS}.
\item Sedimentation is negligible.
\end{enumerate}

\subsection{Governing equations}\label{sec:equations}

In the cylindrical coordinates, free surface of the liquid phase, $z(r,t) $, is described by the equation
\begin{equation*}\label{eq:z}
    z(r,t) = h_f(t) + H(r,t) =  \sqrt{\left(\frac{H_0^2(t)+
    r^2_f(t)}{2 H_0(t)}\right)^2-r^2}
    - \frac{r^2_f - H_0^2(t)- 2 h_f(t)H_0(t)}{2 H_0(t)},
\end{equation*}
where $h_f(t)$ is the height of the phase boundary between sol and gel, $r_f(t)$ is the radial coordinate of the phase front, $H(r,t)$ is the profile of the free surface above liquid (sol) phase and $H_0(t)=H(0,t)$ (see Figure \ref{fig:dropevolution}b).

The conservation of  solute yields
\begin{multline}\label{eq:conserv1}
    \frac{d\Phi}{dt}
    \left( r_f^2 h_f + \frac{H_0^3}{6}+ \frac{H_0 r_f^2}{2}\right) + r_f \frac{dr_f}{dt} ( 2 \Phi h_f +\Phi H_0 -
    2 \Phi_g h_f )+\\ + \frac{\Phi}{2} \frac{dH_0}{dt}(H_0^2 +r_f^2) +\Phi
    r_f^2\frac{dh_f}{dt}=0.
\end{multline}
All quantities, $\Phi(t)$, $r_f(t)$, $h_f(t)$, are the functions of time, $t$. We omit arguments here and below to simplify notation.

The conservation of  solvent yields
\begin{multline}\label{eq:conserv2}
    -\frac{d\Phi}{dt}
    \left( r_f^2 h_f + \frac{H_0^3}{6}+ \frac{H_0}{2} r_f^2\right)+r_f \frac{dr_f}{dt} ( 2 h_f (\Phi_g -\Phi) + H_0(1-\Phi)) +\\ + \frac{(1-\Phi)}{2} \frac{dH_0}{dt}(H_0^2 +r_f^2)
    +(1-\Phi)
    r_f^2\frac{dh_f}{dt}= - \frac{2}{\rho}\int\limits_0^{r_f} J(r,t)\sqrt{1+\left(\frac{\partial
h}{\partial r}\right)^2} r\,dr,
\end{multline}
where $\rho$ is the density of solution.

Sum of \eqref{eq:conserv1} and \eqref{eq:conserv2}
gives the equation for the solution
\begin{equation}\label{eq:conserv}
    r_f H_0 \frac{dr_f}{dt}  + \frac{1}{2} \frac{dH_0}{dt}(H_0^2 +r_f^2)
    + r_f^2\frac{dh_f}{dt}= - \frac{2}{\rho}\int\limits_0 ^{r_f} J(r,t)\sqrt{1+\left(\frac{\partial
h}{\partial r}\right)^2} r\, dr.
\end{equation}

Note that right hand side of \eqref{eq:conserv2}, \eqref{eq:conserv} is proportional to drop mass variation in time
\begin{equation*}\label{eq:massrate}
   \frac{dm}{dt} = -2 \pi \int\limits_0 ^{r_f} J(r,t)\sqrt{1+\left(\frac{\partial
h}{\partial r}\right)^2} r\, dr.
\end{equation*}
We denote
\begin{equation*}\label{eq:defA}
    \frac{2}{\rho}\int\limits_0 ^{r_f} J(r,t)\sqrt{1+\left(\frac{\partial
h}{\partial r}\right)^2} r\, dr = A(t).
\end{equation*}

In spite of the fact that during evaporation, drop mass decrease looks almost linearly~\cite{Annarelli2001,Brutin2011JFM,Sobac2011PRE,Yakhno2010TP}, deviation of the evaporation rate from constant is fairly visible~\cite{Sobac2011PRE}. We suppose that evaporation rate is proportional to the area of free liquid surface and decreases when volume fraction of colloidal particles increases
\begin{equation}\label{eq:A}
A(t) = J_0 r_f^2(t) \left( 1-{\frac {\Phi(t) }{\Phi_g}} \right).
\end{equation}
To our best knowledge, the experiments regarding density of vapor flux above two-phase desiccated sessile drops are not performed yet, so reasonableness of our choice \eqref{eq:A} can be examined only by simulations.

Mass conservation for the solution can be rewritten as
\begin{equation}\label{eq:conservful}
    r_f H_0 \frac{dr_f}{dt}  + \frac{1}{2} \frac{dH_0}{dt}(H_0^2 +r_f^2)
    + r_f^2\frac{dh_f}{dt}= -  A.
\end{equation}

We suppose
\begin{equation}\label{eq:postulat}
    \frac{d}{dt}(h_f +H_0)= V_0\left(1-\frac{\Phi(t)}{\Phi_g}\right),
\end{equation}
where constant $V_0$ is the rate of apex height decrease for a drop of pure solvent. Our assumption is supported by experiments (see, \emph{e.g.}, \cite{Annarelli2001,Sobac2011PRE}).

The set of equations \eqref{eq:conserv1}, \eqref{eq:conservful}, \eqref{eq:postulat} is underdetermined. Additional assumptions are required.

From physical point of view, contact angle between sol, gel, and air, $\Theta$, have to be constant for arbitrary position of the phase boundary, if the volume fraction of colloidal particles inside liquid phase is fixed. Obviously, for the high concentrated colloids ($\Phi\approx\Phi_g$), $\Theta$ have to tend to zero. Contrary, for the dilute colloids ($\Phi \ll \Phi_g$), $\Theta \to \Theta_0$, where $\Theta_0$ is the contact angle between gel, air, and pure solvent. We assume
\begin{equation}\label{eq:thetat}
    \Theta(t) = \Theta_0 \left(1-\frac{\Phi(t)}{\Phi_g}\right).
\end{equation}
In fact, this speculation is valid for the static equilibrium of the phases. If the phase boundary moves, not only surface tension, but viscous forces should be involved into consideration.
Unfortunately, to our best knowledge, there is not any published measurements of $\Theta_0$, thus, it is an adjustable parameter in our model.

\begin{figure}[hptb]
  \centering
  \includegraphics*[width=0.4\textwidth]{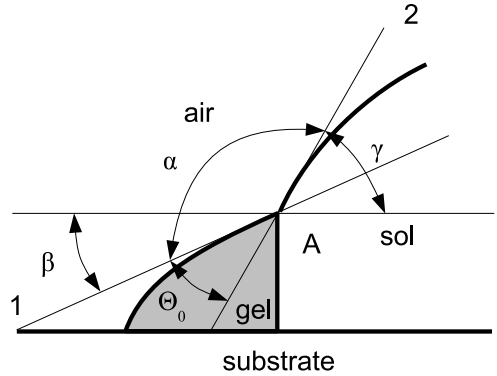}\\
  \caption{1 and  2 are the tangent lines in point B to the free surface of gelled phase and liquid (sol) phase, respectively. $\alpha$ is the angle between these lines, $\beta$ and $\gamma$ are the angles between tangent lines and horizontal direction.}\label{fig:angle}
\end{figure}

At the point A with the coordinates $(r_f,h_f)$, angle $\gamma$ between tangent line to the liquid phase and horizon,  angle $\beta$ between tangent line to the gelled phase and horizon (see Figure
\ref{fig:angle}) can be found from the relation
\begin{equation}\label{eq:tg}
    \tan \gamma(t) = \frac{2 r_f H_0(t)}{r^2_f - H^2(0,t)}.
\end{equation}

It is clear, that
\begin{equation}\label{eq:tg_2}
    \tan \beta(t) = - \frac{dh_f}{dr_f}
\end{equation}
and $\Theta=\pi-\alpha=\pi-(\pi-\gamma+\beta)=\gamma -\beta$ (see Figure \ref{fig:angle}). Hence, from~\eqref{eq:tg} and \eqref{eq:tg_2} yields
\begin{equation}\label{eq:dhdr}
    \frac{dh_f}{dt} \left(1+\tan \Theta(t) \frac{2 r_f H_0(t)}{r^2_f - H^2(0,t)}\right) =
    \frac{dr_f}{dt} \left(\tan \Theta(t) - \frac{2 r_f H_0(t)}{r^2_f -
    H_0^2(t)}\right),
\end{equation}
where $\Theta(t)$ is defined by Equation~\ref{eq:thetat}.

Set of equations \eqref{eq:conserv1}, \eqref{eq:conservful}, \eqref{eq:postulat}, \eqref{eq:thetat}, \eqref{eq:dhdr} is complete and allows calculate the shape of two-phase colloidal drop and volume fraction of colloidal particles inside sol at any time.

It is convenient to measure all lengths in units of initial drop radius (radius of its constant base). In this case $0\leqslant r \leqslant 1$.

\subsection{Parameters of the model}\label{sec:parameters}

It is rather difficult to extract necessary information from the published  experimental data.

To estimate parameters of the model, we use, particularly, the data about desiccation of drops of a phosphate buffered saline solution ($\mathrm{pH} = 7.4$, ionic strength $I = 0.2$ M) of bovine serum albumin (BSA)~\cite{Annarelli2001}. For the concentration of BSA 40 gl$^{-1}$, initial drop mass is about 15 mg, initial height is about 1.5 mm, and drop radius is about 2.5 mm. Thus, initial ratio of height to radius is about 0.6. Central zone of the deposit is flat and its height is $36 \pm 2$ $\mu$m. Thus, ratio of final deposit height to the initial drop height is about 0.024. For the concentration of BSA 60 gl$^{-1}$, central part of the deposit is about 60 $\mu$m, while the rim is about 80 $\mu$m.

For the drop with radius 1.75 mm and concentration of albumen 100 gl$^{-1}$, the heights of the dip and rim are about 15 and 35 $\mu$m, respectively~\cite{Kalyanov2010eng}.

For the drops of native blood plasma and radius 3--3.5 mm, the heights of the dip and rim are about 10 and 20--40 $\mu$m, respectively~\cite{Lychagov2009eng}.

In our computation, initial volume fraction of colloidal particles $\Phi_0$ varies from 0.04 to 0.4, and ratio of initial drop height, $H(0,0)$ to its radius varies from 0.06 to 0.6.

\vspace*{0.5cm}

\section{Results}\label{sec:results}

\begin{figure}[htbp]
\includegraphics[width=0.5\textwidth]{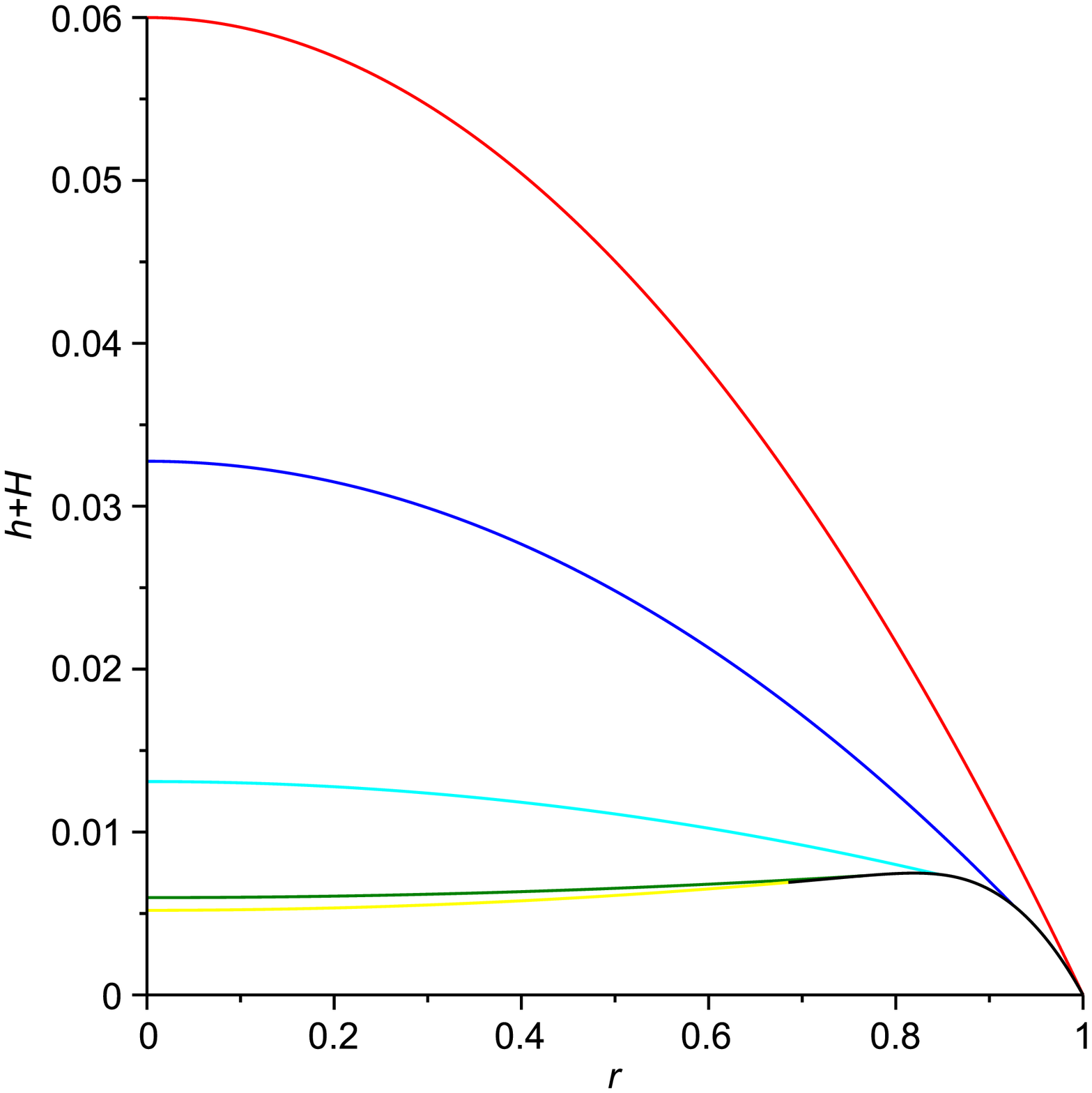}
\includegraphics[width=0.5\textwidth]{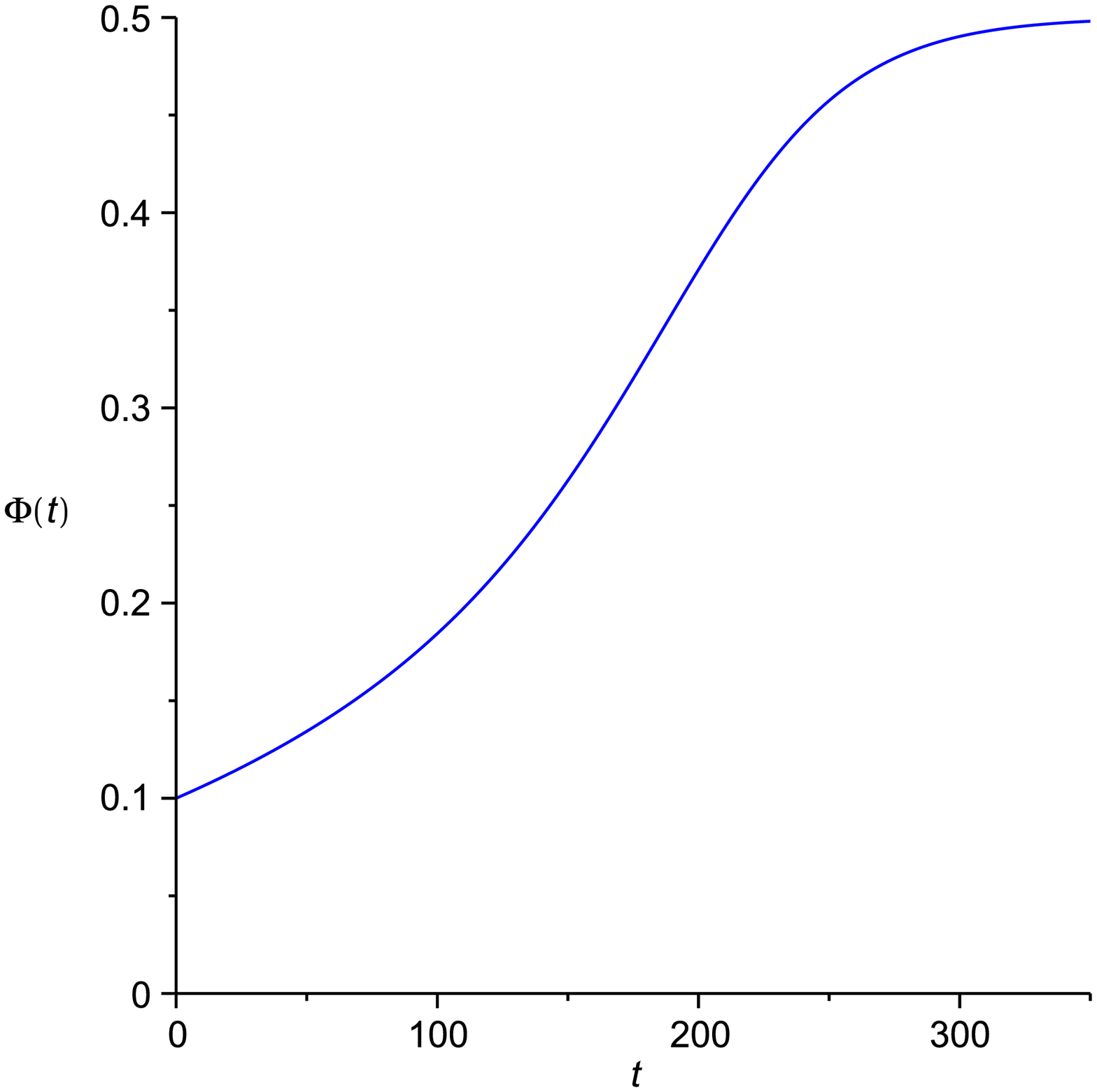}\\
\includegraphics[width=0.5\textwidth]{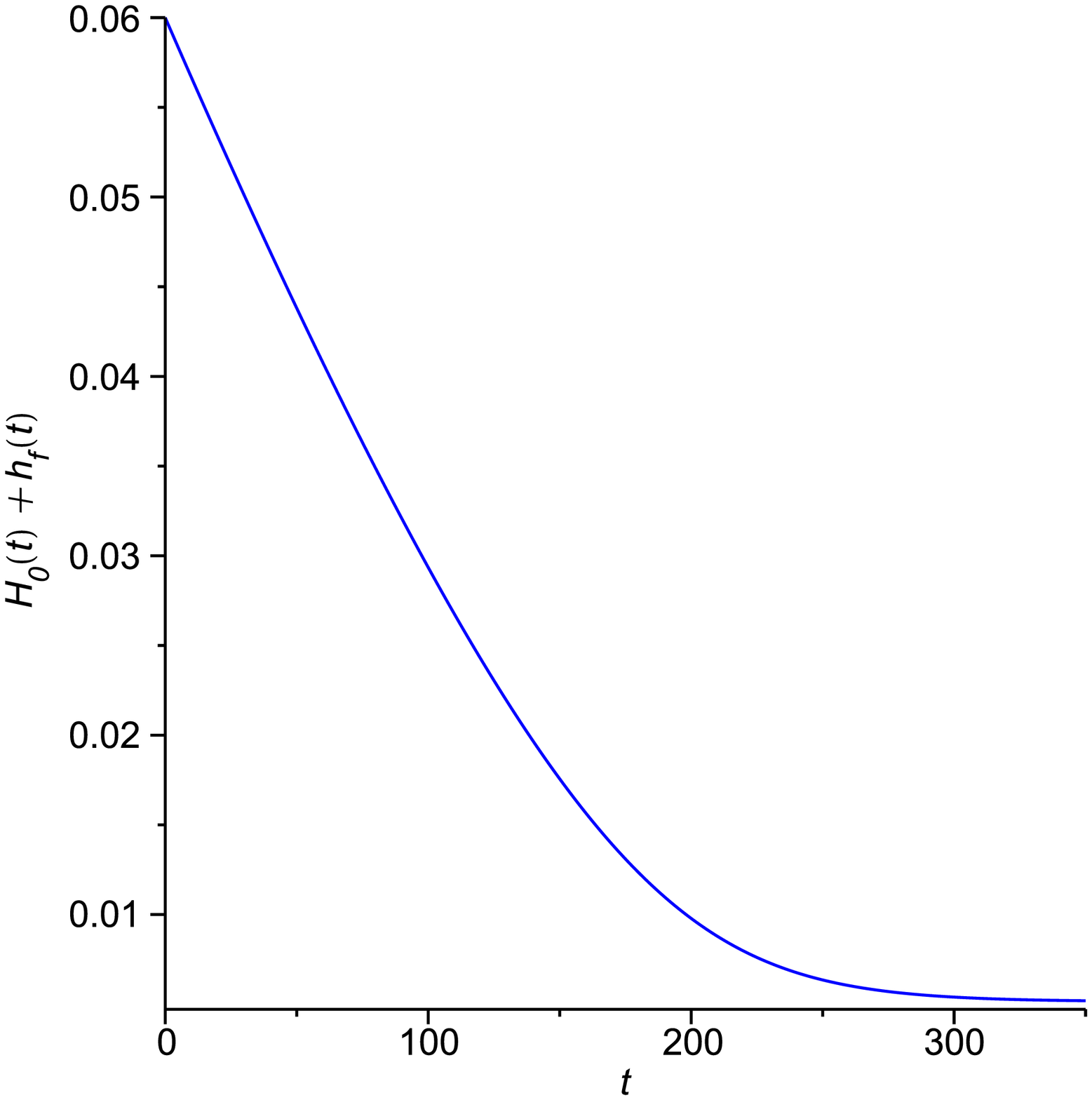}
\includegraphics[width=0.5\textwidth]{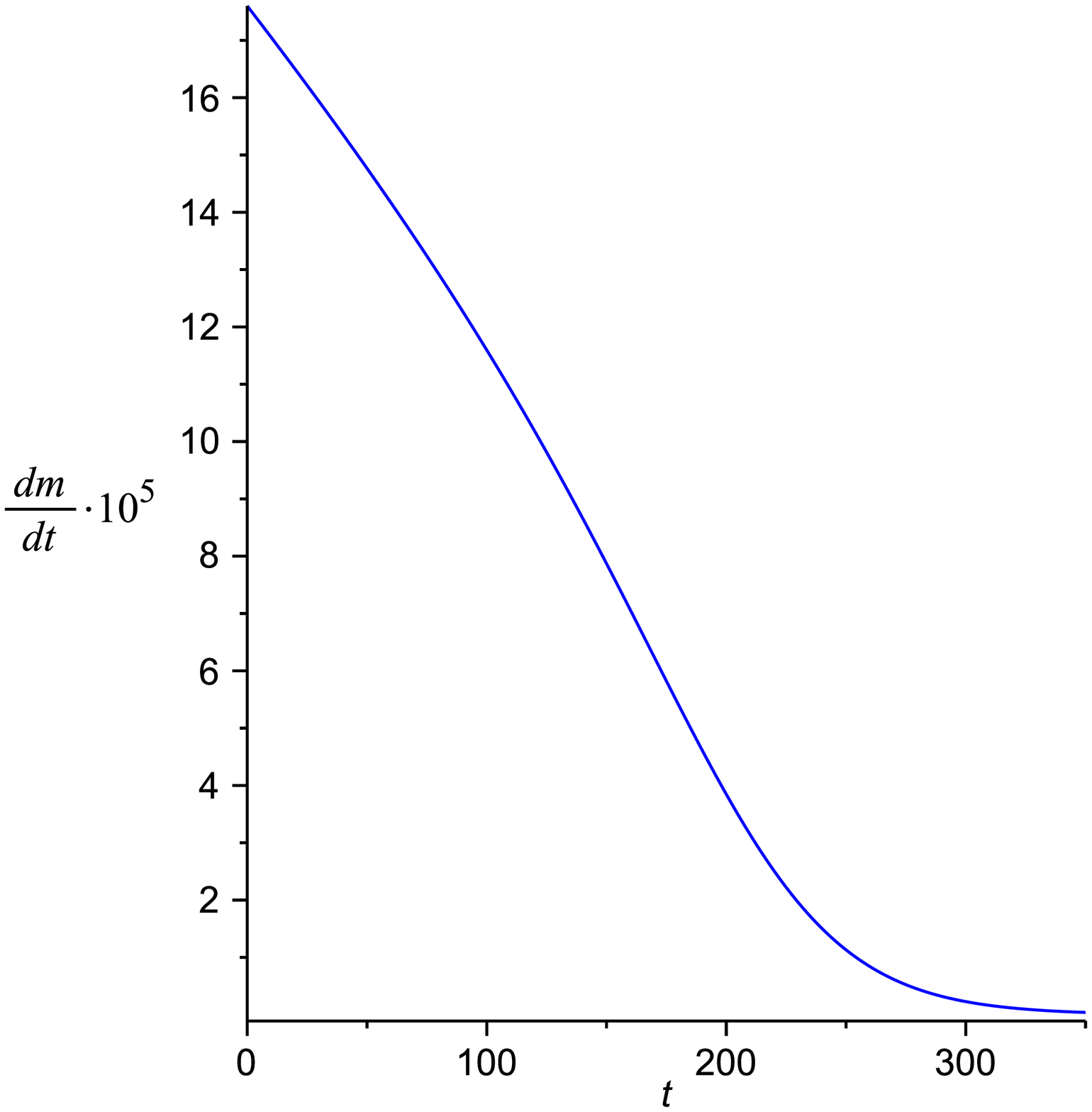}
\caption{Profile, volume fraction, apex, mass lost per time. Thin drop.}
 \label{fig:results}
\end{figure}

Figure \ref{fig:results} demonstrates our simulations for $H(0,0)/R=0.06$, $\Phi(0)=0.1$. Evaporation rate is $J_0 = 2.2\cdot 10^{-4}$, velocity of apex decreasing is $V_0 =4.22 \cdot 10^{-4}$, and adjustable parameter is $\Theta_0 = \frac {\pi}{160}$. Different profiles correspond (from up to down) to the initial time ($t=0$), $t=t_D/4$, $t=t_D/2$, $t=3t_D/4$, and $t=t_D$. Final profile of the gelled film is close to the experimentally observed \cite{Annarelli2001,Kalyanov2010eng,Lychagov2009eng}.


Dynamics of volume fraction of the colloidal particles looks like reported by Sobac and Brutin in \cite{Sobac2011PRE}. The experiment \cite{Sobac2011PRE} shows rather sharp decreasing of the drying rate, $|dm/dt|$, when the system changes from two-phase (sol and gel) to one-phase (gel). Unfortunately, our computations of drying rate don't demonstrate the same excellent agreement with the experiment, \emph{i.e.} drying rate decreases monotonically. We suppose the results of simulations may be improved by means of more accurate definition of \eqref{eq:A}.

\begin{figure}[htbp]
\centerline{\includegraphics[width=0.5\textwidth]{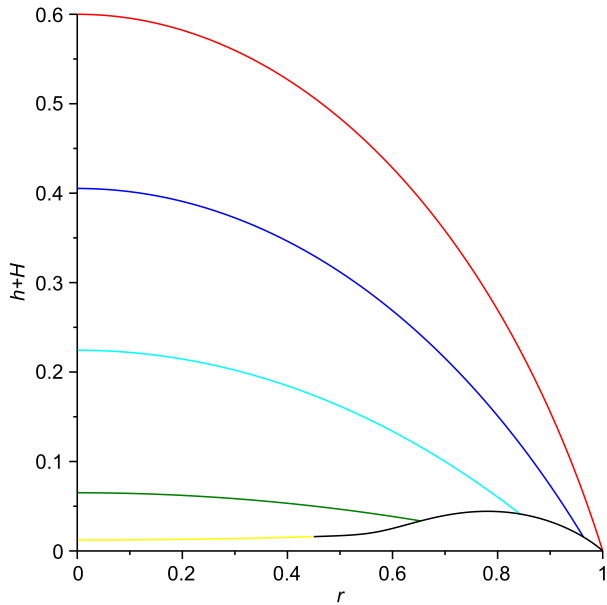}}
\caption{Dynamics of shape. Thick drop.}
 \label{fig:annarelli}
\end{figure}

Figure \ref{fig:annarelli} shows evolution of the drop during desiccation if initial volume fraction of the suspended colloidal particles is 0.04 and the drop is rather thick ($H(0,0)/R=0.6$). These parameters correspond to the experiments by Annarelli \emph{et al.} \cite{Annarelli2001}. Adjustable parameter $\Theta_0$ is $\pi/8$. Instants are the same as in Figure  \ref{fig:results}.

\vspace*{0.5cm}

\section{Conclusion}\label{sec:conclusion}

Proposed semiempirical model give simple explanation of the colloidal sessile drop desiccation. The results obtained for the parameters corresponding blood serum are consistent with experiments. The model allows to describe correctly the final shape of the dried film (deposit).

\vspace*{0.5cm}

\section*{Acknowledgements}
The authors are grateful to the Russian Foundation for Basic Research for the financial support (grant 09-08-97010-r\_povolzhje\_a).



\end{document}